\documentclass[conference,a4paper]{IEEEtran}
\usepackage{xcolor}
\usepackage{balance}
\usepackage{amsmath,bm,amsfonts,amssymb,graphicx,color,multirow,setspace,xfrac,comment,cite}
\begin{document}
%
\title{Impact of Spatially Consistent Channels on Digital Beamforming for Millimeter-Wave Systems\vspace{-2pt}}

\author{\IEEEauthorblockN{
Harsh Tataria and      
Fredrik Tufvesson      
}                      
\IEEEauthorblockA{
Department of Electrical and Information Technology, 
Lund University, Lund, Sweden}
 \IEEEauthorblockA{e-mail: \{harsh.tataria, fredrik.tufvesson\}@eit.lth.se}
 \vspace{-23pt}
}


\IEEEspecialpapernotice{(Invited Paper)\vspace{-2pt}}

\maketitle

\begin{abstract}
The premise of massive multiple-input multiple-output (MIMO) is based around coherent transmission and detection. Majority of the vast literature on massive MIMO presents performance evaluations over simplified statistical propagation models. All such models are drop-based and do not ensure continuity of channel parameters. In this paper, we quantify the impact of spatially consistent (SC) models on beamforming for massive MIMO systems. We focus on the downlink of a 28 GHz multiuser urban microcellular scenario. Using the recently standardized Third  Generation Partnership Project 38.901 SC-I procedure, we evaluate the signal-to-interference-plus-noise ratio of a user equipment and the system ergodic sum spectral efficiency with zero-forcing, block diagonalization, and signal-to-leakage-plus-noise ratio beamforming. Across different user movement trajectories, our results disclose that SC channels yield a significant performance loss relative to the case without SC due to substantial spatial correlation across the channel parameters. 
\vspace{-10pt}
\end{abstract}

\begin{IEEEkeywords}
Beamforming, ergodic sum spectral efficiency, massive MIMO, mmWave, SINR, spatial consistency. 
\end{IEEEkeywords}

%

\vspace{-13pt}
\section{Introduction}
\label{Introduction}
\vspace{-7pt}
Significant advancements in semiconductor technologies, and the potential to offer large radio frequency (RF) bandwidths makes millimeter-wave (mmWave) frequencies an important part of fifth-generation (5G) cellular systems \cite{SHAFI2}. Relative to systems below 6 GHz, mmWave systems face a unique set of challenges, which constitute the major limiting factors towards their large-scale rollout. Firstly, since free-space attenuation increases quadratically with frequency (for frequency independent antenna gain), a large array gain is required to penetrate the transmitted waveform to moderate distances. This makes the use of massive antenna arrays at the base station (BS) essential at mmWaves. This is unlike sub-6 GHz systems, where they are beneficial and are primarily used for coherent transmission/reception of signals. Furthermore, the efficiency of diffraction strongly decreases since common objects throw sharp shadows. Supporting the theoretical speculations, most current mmWave measurements demonstrate that the nature of the channel is sparse and directional (see e.g., \cite{KO1,NIST1}, and references therein for a taxonomy).

Since vast majority of the literature on mmWave channels focuses on scenarios with fixed BSs and user equipments (UEs), there is a general lack of understanding into the behavior of small and large-scale parameters in dynamic scenarios. In reality, UEs can not be assumed to be completely stationary, since even very small movements can contribute to large phase variations of the contributing multipath components (MPCs). To this end, the channel parameters need to be continuously evolving as the UE moves into and out of regions that are covered by the BS with different propagation mechanisms. The authors of \cite{BAS1} present measured investigations into the dynamic behavior of mmWave channels characterizing the channel's delay and angular parameters. For system-level evaluation of performance, in order to model continuously evolving channels, the inclusion of spatial consistency (SC) is mandatory. SC ensures that UEs experience a similar scattering environment causing smoother channel transitions with relative motion. This is in contrast to the so-called drop-based models, in which channel segments and parameters between multiple UE ``drops" do not have continuity. This has been lately recognized by the Third Generation Partnership Project (3GPP), who have standardized SC modelling in Release 14 TR 38.901 \cite{3GPPTR38901}. Specifically, two SC procedures are proposed for modelling the continuity of channel parameters over a coherence distance of 15 m. Based on this, the authors of \cite{JU1} adapted the New York University Simulator model to have a SC procedure over UE moving distances of 15 m. Further work on evaluating the 3GPP SC procedures has been reported in \cite{KURRAS1,ADEMAJ1} using the Quasi Deterministic Radio Channel Generator model at 2 GHz. 

In parallel to the advancements in dynamic mmWave propagation models, progress has also been made in the design of the BS transceiver architectures. Both fully digital and analog-digital (a.k.a. hybrid) beamforming structures have been proposed at mmWaves, and their implementation aspects have been the topic of long standing debates in the literature \cite{YANG1,HU1}. The main motivation behind the hybrid beamforming architecture is to reduce the net energy consumption of mixed signal components by lowering the number of active RF chains with the use of explicit beamforming. While this improves the energy efficiency of mmWave systems, the spectral efficiency performance is lower than the fully digital case due to the loss in the effective degrees-of-freedom. On the other hand, fully digital architectures with implicit beamforming also show promise and have been practically realized in \cite{YANG1,HU1} at 28 GHz for BS arrays up to 64 antennas. Despite this, for both architecture types, majority of the proposed designs focus on optimizing the ergodic spectral efficiency of the system with fixed BS and UEs using the drop-based models (see e.g., \cite{SHAFI2,YANG1,HU1,LARSSON1,TATARIA1}). \emph{To the best of our knowledge, the fundamental impact of UE mobility, and hence SC, on dynamic beamforming remains unexplored. This is critical, since majority of the evaluations inaccurately estimate the resulting performance without considering SC. Unlike previously, in this paper, we quantify the performance  of common digital  beamforming methods (described later) in SC mmWave channels.}

\textbf{Contributions.} On the downlink of a 28 GHz urban microcellular (UMi) scenario, we consider UEs moving along different trajectories, and model SC with the 3GPP SC-I procedure described in \cite{3GPPTR38901}\footnote{Due to space limitations, we only present one such scenario in the paper.}. In doing so, we present the correlated evolution of MPC parameters across the UE routes. To quantify the impact of SC on multiuser beamforming performance, with a 16$\times$16 cross-polarized uniform planar array (UPA) at the BS, we evaluate the instantaneous signal-to-interference-plus-noise ratio (SINR) of each UE having 4$\times$4 UPAs with zero-forcing (ZF), block diagonalization (BD) and signal-to-leakage-plus-noise ratio (SLNR) beamforming. The SINR evaluations serve as a means of predicting the ergodic sum spectral efficiency over all UE routes. With ZF, BD, and SLNR processing at the BS, our results demonstrate that SC yields a significant performance loss relative to the case without SC due to significant correlation in the channel parameters. The presented results can be interpreted as a cautionary tale towards the impact of beamforming in dynamic channels relative to channels which are drop-based. 

\begin{figure}[!t]
\begin{center}
\includegraphics[width=7.3cm]{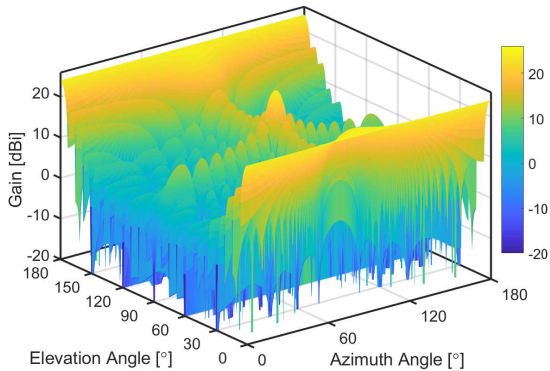}
\vspace{-13pt}
\caption{BS array pattern as a function of azimuth and elevation scan angles.}
\end{center}
\label{BSArrayPattern1}
\vspace{-11pt}
\end{figure}
\begin{figure}[!t]
\vspace{-9pt}
\begin{center}
\includegraphics[width=7.3cm]{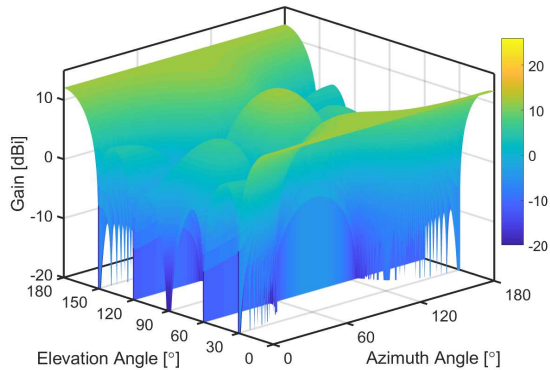}
\vspace{-10pt}
\caption{UE array pattern as a function of azimuth and elevation scan angles.}
\end{center}
\vspace{-20pt}
\label{UEArrayPattern1}
\end{figure}

\vspace{-4pt}
\section{System, Channel and Signal Models}
\label{SystemPropagationChannelandSignalModels}
\vspace{-1pt}
We consider a cellular BS operating at 28 GHz in an UMi environment simultaneously serving $L$ UEs within the same time-frequency resource. We assume that the total number of UEs are much less than the total number of antenna elements at the BS. The BS comprises of a UPA on the $x-z$ plane with $M=256$ co-located antenna elements. The BS elements are configured in eight rows and 16 columns with \emph{two} polarization states. Each antenna element has a slant angle of $\pm{}45^\circ$. The electrical distance between adjacent co-polarized elements is $\lambda/2$ in the azimuth and $\lambda$ in the elevation domains, where $\lambda$ denotes the carrier wavelength. Consistent with \cite{SHAFI2,3GPPTR38901}, the directional gain per-element is 8 dBi, with the horizontal and vertical patterns as stated in Table 7.3-1 on page 22 of \cite{3GPPTR38901}. The BS array pattern across different azimuth and elevation angles with the aforementioned per-element pattern can be depicted in Fig.~1. From the figure, one observe that with 256 elements, a half-power beamwidth (HPBW) of $8^\circ$ in both azimuth and elevation is achieved in the broadside direction of the UPA. Naturally, the array gain is heavily influenced by the azimuthal and elevation scan angle. On the UE side, a 16 element UPA is employed (at each UE) with two rows and four columns, also having two polarization  states. Equivalent inter-element spacing to the BS is assumed in both the azimuth and elevation domains. Figure~2 demonstrates the directional dependence of the UE pattern, where one can observe a much wider HPBW in the azimuth and elevation domains of approximately 
$63^\circ$ and $32^\circ$, with a per-element gain of 9 dBi. We defer the discussion of UE placements till Sec.~III of the paper where we present the considered scenario. Although the exact BS and UE antenna structure and numbers are as detailed above, in order to retain generality in the system model, we denote the total number of antennas at UE $\ell$ by $N_\ell$, while the total number of elements at the BS is denoted by $M$. We assume uniform power allocation to each UE from the BS, and perfect knowledge of the propagation channel is assumed at both link ends, since the central focus of our work is to quantify on the impact of SC channels on digital beamforming. 

Keeping the above in mind, the $N_\ell\times{}M$ downlink propagation channel from the BS to UE $\ell$ is denoted by $\mathbf{H}_\ell$, and follows the 3GPP cluster-based definition of the mmWave propagation channel from \cite{3GPPTR38901}. We adapt the large-scale (free-space attenuation, shadow fading, probability of line-of-sight, Ricean $K$-factors, and angular spreads) and small-scale (delays, MPC angles-of-departure and angles-of-arrival) parameters for the UMi scenario from Sec. 7.4 of \cite{3GPPTR38901}. For ease of presentation, we refer the interested reader to the discussions from page 24 onward in \cite{3GPPTR38901} for the precise details on the channel impulse response generation. Note that the impulse response includes the directional nature both BS and UE antenna arrays. The received signal at UE $\ell$ is 
\vspace{-2pt}
\begin{equation}
    \label{RXSignalatUEl}
    \mathbf{r}_\ell=\mathbf{H}_\ell\hspace{1pt}
    \mathbf{s}+\mathbf{n}_{\ell}, 
    \vspace{-2pt}
\end{equation}
where $\mathbf{s}$ is a $M\times{}1$ complex input signal and $\mathbf{n}_{\ell}$ is the $N_{\ell}\times{}1$ additive white Gaussian noise vector at the input of 
UE $\ell$. More specifically, each entry of 
$\mathbf{n}_{\ell}\sim\mathcal{CN}\left(\hspace{1pt}0,\sigma^2\right)$. For simultaneous service to each UE, we consider the use of more simpler \emph{linear} digital beamforming approaches as done extensively for the canonical form of massive MIMO below 6 GHz \cite{LARSSON1,TATARIA1}. With the above, the composite transmitted signal with linear beamforming can be written as 
\vspace{-6pt}
\begin{equation}
\label{TXSignalWithBeamforming}
    \mathbf{s}=\mathbf{Xd}=
    \sum\limits_{\ell=1}^{L}
    \hspace{2pt}\mathbf{X}_{\ell}
    \hspace{2pt}\mathbf{d}_{\ell},
    \vspace{-3pt}
\end{equation}
where $\mathbf{X}$ is the $M\times{}m$ beamforming matrix and $\mathbf{d}$ is the data bearing vector of transmitted symbols which has the dimension $m\times{}1$. Furthermore, $\mathbf{X}_{\ell}$ is the beamforming matrix for UE $\ell$ with dimensions $M\times{}m_\ell$. To this end, $\mathbf{d}_\ell$ is the desired data for UE $\ell$ and is of size $m_\ell\times{1}$ for $m_\ell$ symbol streams to UE $\ell$. By virtue of this, we have $m=\sum\nolimits_{\ell{}=1}^{L}m_\ell$. With this model, the received signal at UE $\ell$ is given by 
\vspace{-4pt}
\begin{equation}
    \label{RXSignalUEl3}
    \mathbf{r}_{\ell}=\hspace{3pt}
    \mathbf{H}_{\ell}\hspace{1pt}
    \mathbf{X}\mathbf{d}+\mathbf{n}_{\ell}
    =\mathbf{H}_{\ell}\hspace{3pt}
    \mathbf{X}_{\ell}
    \hspace{3pt}\mathbf{d}_{\ell}+\textbf{H}_{\ell}
    \sum\limits_{\substack{k=1\\k\neq{}\ell}}^{L}
    \mathbf{X}_{k}\hspace{2pt}\mathbf{d}_{k}+
    \mathbf{n}_{\ell},  
    \vspace{-2pt}
\end{equation}
where the first term denotes the desired received signal, the second term represents the multiuser interference and the third term denotes the additive 
white Gaussian noise. The expression in \eqref{RXSignalUEl3} can be translated into a SINR for UE $\ell$ as 
\vspace{1pt}
\begin{equation}
    \label{SINRUEl}
    \textrm{SINR}_\ell=
    \frac{m_\ell^{-1}\hspace{1pt}
    \mathbb{E}\left[\left\|\hspace{1pt}\mathbf{H}_\ell\hspace{1pt}
    \mathbf{X}_\ell\hspace{2pt}\mathbf{d}_\ell\right\|^2\right]}
    {\mathbb{E}\left[\hspace{1pt}\left\|\hspace{1pt}\mathbf{H}_\ell
    \sum\nolimits_{\substack{k=1\\k\neq{}\ell}}^{L}
    \mathbf{X}_k\hspace{1pt}\mathbf{d}_k+\mathbf{n}_\ell
    \hspace{1pt}\right\|^2\right]}, 
    \vspace{1pt}
\end{equation}
where the expectation is performed over the ensemble of 
$\mathbf{d}_\ell$ and $\mathbf{d}_k$, respectively. 
In a similar line, the average transmit power at the BS is defined by
\begin{equation}
    \label{TotalTXPoweratBS}
    \mathbb{E}\left[\mathbf{d}^{H}\mathbf{X}^H\mathbf{X}
    \mathbf{d}\right]=
    \phi^2\left[
    \hspace{1pt}\textrm{Tr}
    \left(\mathbf{X}^{\hspace{-1pt}H}
    \mathbf{X}\right)\hspace{1pt}\right],
\end{equation}
where the expected value is over 
the data components of $\mathbf{d}$. As such, 
$\phi^2\hspace{-1pt}=\hspace{-1pt}\mathbb{E}\hspace{1pt}[\hspace{1pt}|\hspace{1pt}(\mathbf{d}_\ell)_r|^2\hspace{1pt}]$, where $(\mathbf{d}_\ell)_r$ denotes 
the $r$-th element of $\mathbf{d}_\ell$. For simplicity, we assume that all streams have equal power. Though we do not consider the extension to unequal power per-stream in this work, we note that it is possible to investigate this from the current model framework. Without loss of generality, we let 
$\phi^2=1$, and normalize the global 
beamforming matrix, $\mathbf{X}$, such that $\textrm{Tr}\hspace{1pt}(\mathbf{X}^{\hspace{-1pt}H}
\hspace{-1pt}\mathbf{X})=1$. Considering this, the average transmit power is normalized to unity with the operating SNR being $1/\sigma^2$. With equal power per-UE, the transmit SNR can be written as $(\sigma^2{}L)^{-1}$. The SINR in \eqref{SINRUEl} for the $\ell$-th UE can be translated into the ergodic sum spectral efficiency (in bits/s/Hz) over all $L$ UEs via 
\begin{equation}
    \label{ErgodicSumSpectralEfficiency}
    \textrm{R}_{\textrm{sum}}=\mathbb{E}\left[
    \sum\limits_{\hspace{2pt}
    \ell=1}^{L}\log_2\left(1+\textrm{SINR}_\ell\right)\hspace{1pt}\right]. 
\end{equation}
Here the expectation is taken over the small-scale fading in $\mathbf{H}_\ell$, where each realization of $\mathbf{H}_\ell$ is generated with a particular azimuth and elevation power-angular statistics as defined in \cite{3GPPTR38901} for the UMi scenario. We use the SINR and ergodic sum spectral efficiency as the key performance measures to evaluate the impact of SC on multiuser beamforming techniques. 

In what follows, we use ZF, BD and SLNR beamforming 
to design $\mathbf{X}$. Due to space limitations, we omit providing the 
exact mathematical expressions for the design of $\mathbf{X}$ considering 
the aforementioned techniques. Instead, we refer the interested readers 
to \cite{SPENCER1,TATARIA1,SADEK1,LARSSON1} for more 
detailed reading. In Sec.~IV of the paper, 
we evaluate the performance of ZF, BD and SLNR beamforming techniques 
with dynamic SC channels. In the sequel, we discuss the dynamic scenarios, 
SC modelling methodology of 3GPP, and present the corresponding 
implications on the propagation parameters.

\vspace{-5pt}
\section{SC Modelling and Implications on Propagation Parameters}
\label{SpatialConsModellingImplicationsonChannel}
In line with the descriptions in Sec.~\ref{SystemPropagationChannelandSignalModels}, for evaluating the impact of SC channels on mmWave beamforming, we consider the scenario presented in Fig.~3. The BS (marked with a blue cross) is located approximately 20 m away from UE 1 (blue square) and 40 m from UE 2 (red square). Both UEs are simultaneously moving from their initial positions depicted in Fig.~3. The trajectory of UE 1 finishes at UE 2's initial position, while the UE 2's trajectory is completed at the end of the simulated route shown in Fig.~3 with the black dotted line. The UEs move at the interval of 0.1 with a velocity of 0.83 m/s. The propagation channel is modelled as in 3GPP TR 38.901 \cite{3GPPTR38901}, where six scattering clusters are assumed between the BS and UEs. The specific cluster characteristics in both azimuth and elevation domains follow the 3GPP defined statistical distributions. The update distance of UEs is set to 0.1 m, so that we are able to accurately capture small variations in the propagation channel parameters in comparison to the 15 m correlation distance of large-scale parameters defined by the 3GPP \cite{3GPPTR38901}. Correlation distance in the 3GPP is defined as the distance beyond which the autocorrelation value of a large-scale parameters drops below 0.5. It is important to note that the correlation distance varies according to different large-scale parameters such as angular spread, delay spread, Ricean $K$-factor and shadow fading. For all of these parameters, in an UMi mmWave environment, correlation distances typically tend to vary from 12 to 15 m \cite{3GPPTR38901}. On this line, the authors of \cite{RAPPAPORT2} also observed high correlation across the average received power levels over grids with 5 m lengths, even though directional antenna elements showed the small-scale correlation of received power was less than one meter, and  heavily dependent on the element look angles. The employed 3GPP-based SC modelling procedure is highlighted later in the text of the paper. Since majority of modelling methodology is reported in \cite{3GPPTR38901}, we only provide highlights of the approach employed by the 3GPP. We note that an implicit assumption during the subsequent evaluations is that the BS is constantly able to track/steer the gain in the direction of the UE's mainlobe, in order to maintain continuous communication over the entire UE trajectories. 
\begin{figure}[!t]
\vspace{-10pt}
\begin{center}
\includegraphics[width=7.3cm]{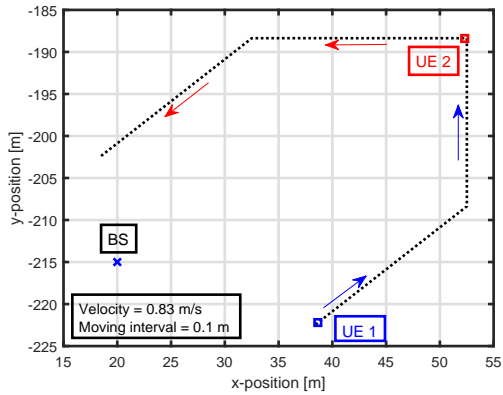}
\vspace{-5pt}
\caption{Investigated scenario where both UEs move on different trajectories as indicated by the red and blue arrows.}
\end{center}
\label{ScenarioA}
\vspace{-17pt}
\end{figure}
\begin{figure}[!t]
\begin{center}
\includegraphics[width=7.3cm]{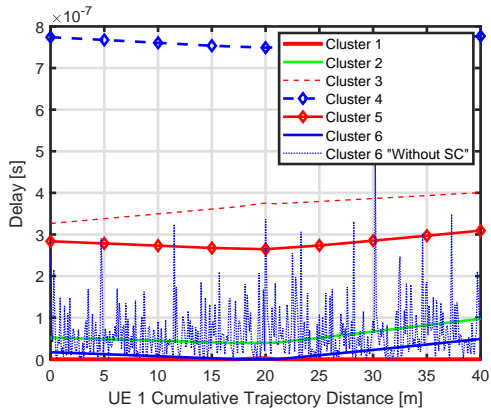}
\vspace{-7pt}
\caption{Cluster delays with and without SC vs. cumulative distance of UE 1's trajectory.}
\end{center}
\vspace{-28pt}
\end{figure}

As mentioned in the introduction of the paper, the 3GPP introduce two procedures for SC modelling for system level simulations. The first approach, known as SC-I, updates the propagation channel cluster powers, delays and angular parameters at time instant $t$ based on the powers, delays, angles, UE velocity vector, and UE position vector at the previous time instant, time $t-\Delta{}t$, following the equations in \cite{3GPPTR38901}. The initial spatially consistent delays, powers and angles of clusters are generated according to the same procedure as without SC. In SC-I, assuming a UE moves along a trajectory at a velocity $v$, its moving interval is then limited based on this, and is within the range of 1 m over a short time epoch $\Delta{}t$. Then, for each $\Delta{}t$ time lapse, the delays, powers, and angles are updated as in \cite{3GPPTR38901}. Finally, the updated parameters are used to generate the SC channel impulse response. Overall, SC-I acts as an iterative algorithm which correlates and updates the channel parameters at each moving distance of the UE. Different to SC-I, in the second approach known as SC-II, small and large-scale fading propagation parameters are still independent for different UE positions, but rather the generation procedure of these parameters is modified. More specifically, the modified steps in \cite{3GPPTR38901} generate delays and angles based on uniform distributions where the coefficients linearly depend on the correlation distance to assure SC of the simulated channel over the UE trajectory. For the purpose of our evaluations, we use SC-I method to generate SC channel impulse responses at each UE position within the trajectories of the considered scenario. 
\vspace{-12pt}
\begin{figure}[!t]
\vspace{-5pt}
\begin{center}
\vspace{-5pt}
\includegraphics[width=7.3cm]{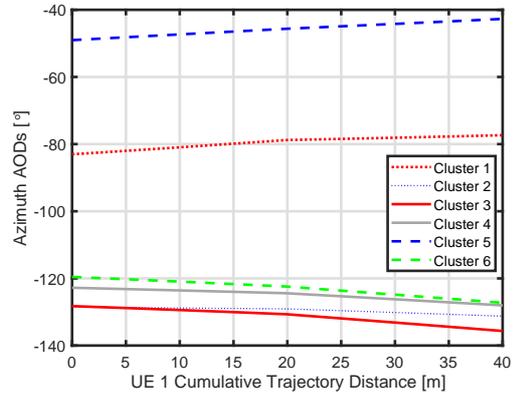}
\vspace{-5pt}
\caption{Azimuth AODs with SC vs. cumulative distance of UE 1's trajectory.}
\end{center}
\vspace{-10pt}
\end{figure}
\begin{figure}[!t]
\vspace{-4pt}
\begin{center}
\includegraphics[width=7.3cm]{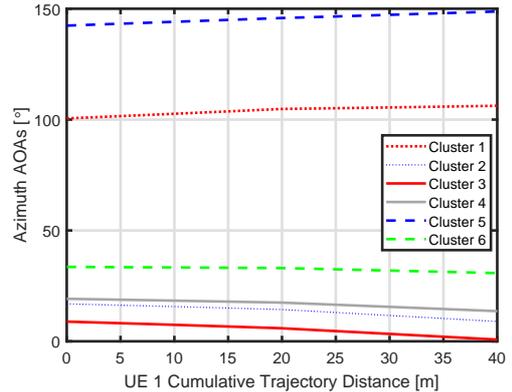}
\caption{Azimuth AOAs with SC vs. cumulative distance of UE 1's trajectory.}
\end{center}
\vspace{-21pt}
\end{figure}

In order to maintain clarity and to minimize cluttering the results, we focus on a subset of propagation parameters, since the conclusions drawn below are also valid for other parameters. The spatially consistent delays of the six scattering clusters are depicted in Fig. 4. From the result, it can be readily observed that SC makes the cluster delays evolve continuously and smoothly through the entire UE track. For the sake of example, UE 1's track is analyzed. In order to analyze the difference between spatially consistent vs. spatially inconsistent delays, we take the example of cluster 6, where the delay response without SC is shown. Here, one can notice a clear difference as the delays tend to fluctuate abruptly and are discontinuous from one channel segment to the next. Similar effects can be observed on the azimuth angles-of-departure (AODs) and arrival (AOAs), which are demonstrated in Figs. 5 and 6, respectively. Here one can also observe the continuous evolution of the cluster angles relative to distance induced by the SC-I procedure. Furthermore, in both figures, the turning point of the UE can be clearly identified with the \emph{change} in the angular parameters at approximately 20 m (half-way point). 

\vspace{-5pt}
\section{Impact of SC on Multiuser Beamforming}
\label{ImpactofSConMultiuserBeamforming}
\vspace{-1pt}
We now consider the impact of SC on the multiuser beamforming techniques. We evaluate the instantaneous SINR and the ergodic sum spectral efficiency performance of the system using the expressions in (4) and (6), respectively. Figure 7 demonstrates the 
cumulative distribution functions (CDFs) of the SINR for UE 1 with an operating SNR of 5 dB. Two trends can be observed: (1) Irrespective of the beamforming technique, spatially consistent channel yield a remarkable 55\% loss in the per-UE SINR relative to drop-based modelling without SC. This is a result of the fact that SC heavily correlates the UE channels at each moving interval. Despite UE 1 being at least 20 m away from UE 2, complete cancellation of multiuser interference does not happen due to the high spatial correlation levels, not allowing for multiuser separation to happen effectively. In stark contrast to this, without SC, since the large-scale parameters vary arbitrarily without any relation to the UE movements, they help to effectively decorrelate the channel, adding diversity to boost performance. (2) The peak SINR levels at CDF = 0.95 exhibit larger variations. This is because at the upper end of the CDFs, more favorable channel conditions occur with lower correlation levels, and hence occasionally unexpectedly high SINRs can be also be observed. In contrast, at the median CDF levels, i.e, around CDF = 0.5, a more clear difference can be observed amongst the three beamforming techniques, where BD is seen to perform best with and without SC. This is unlike ZF and SLNR, which suffer from noise inflation and poor signal power maximization in correlated channels. \begin{figure}[!t]
\begin{center}
\includegraphics[width=8cm]{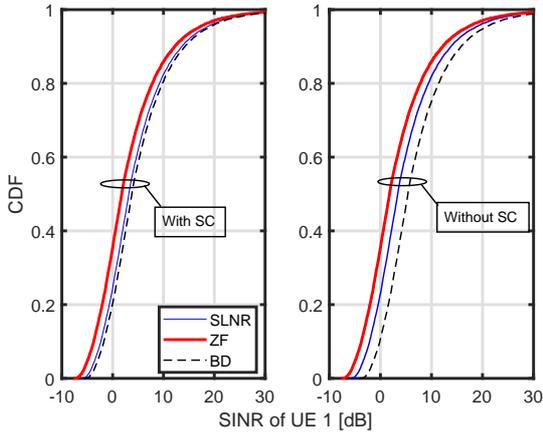}
\vspace{-5pt}
\caption{SINR CDFs of UE 1 with and without SC considering all beamforming techniques. Each point in the CDF is plotted at SNR=5 dB, as the UE travels on its track.}
\end{center}
\vspace{-23pt}
\end{figure}

The combined performance of the UEs 1 and 2 can be evaluated by examining the ergodic sum spectral efficiency. Figure 8 depicts this as a function of the operating SNR level, where it can be seen that the true difference in performance with and without SC is seen at SNR levels beyond 0 dB, since this is when the multiuser system can leverage the multiplexing gains. At SNR = 10 dB for instance, an approximate 60\% degradation in the ergodic sum spectral efficiency is observed with SC in contrast to no SC. 
\begin{figure}[!t]
\vspace{-2pt}
\begin{center}
\hspace{-10pt}
\includegraphics[width=8cm]{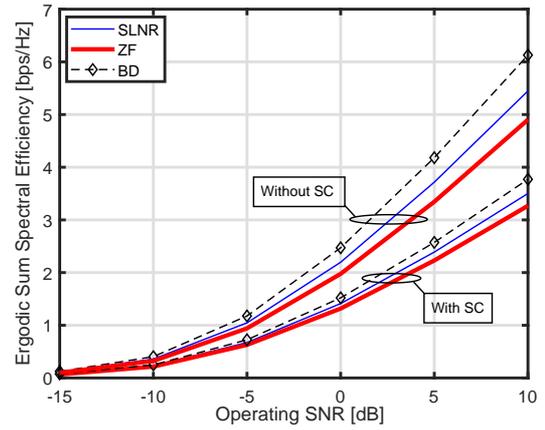}
\vspace{-3pt}
\caption{Ergodic sum spectral efficiency vs. operating SNR with and without SC considering all beamforming techniques.}
\end{center}
\vspace{-16pt}
\end{figure}

\vspace{-8pt}
\section{Conclusions}
\label{Conclusions}
\vspace{-4pt}
Unlike previous studies, this paper quantifies the impact of 
SC on dynamic beamforming in a 28 GHz multiuser UMi scenario. Following the 3GPP TR 38.901 channel model and SC procedure, the instantaneous SINR 
of a UE, as well as the ergodic sum spectral efficiency of the system 
are evaluated with ZF, BD, and SLNR beamforming techniques. A particular mobility scenario is studied as an example, where each UE has its own moving trajectory and velocities. The impact of SC is highlighted firstly on the propagation parameters, followed by an evaluation into the multiuser beamforming performance. Our results demonstrate that drop-based modelling of the channels results in inaccurate performance estimates, and the inclusion of SC yields a large performance loss as the  channels become heavily correlated. This is since the channel parameters continuously evolve and the UE experiences similar scattering. The presented results can be interpreted as a cautionary tale towards the impact of beamforming in dynamic channels relative to channels which are drop-based.

\vspace{-6pt}
\bibliographystyle{IEEEtran}

\end{document}